\documentclass[aps,prd,amssymb]{revtex4}
\usepackage{graphicx} 
\usepackage{dcolumn} 
\usepackage{bm} 
\usepackage{hyperref} 
\usepackage[tbtags]{amsmath} 
\usepackage{xcolor}
\usepackage{palatino}

\begin{document}


\title{Thermodynamics of Regular Black Holes with Cosmic Strings}

\author{Ceren H. Bayraktar}
\affiliation{Department of Physics, \.{I}zmir Institute of Technology, TR35430 \.{I}zmir, Turkey}
\email{cerenbayraktar@std.iyte.edu.tr}

\begin{abstract}

In this article, the thermodynamics of regular black holes with a cosmic string passing through it is studied. We will observe that the string has no effect on the temperature as well as the relation between entropy S and horizon area A.
\end{abstract}

\date{\today}

\maketitle

\section{Introduction}
\paragraph*{}
 The notion of a black hole was firstly predicted by John Michell in 1783. Later, in 1915, the modern concept of black holes was developed with Einstein's general theory of relativity. However, because of the singularity at the center, they are also seen as the proof of the breakdown of the very same theory. We can solve this singularity problem by constructing a spherically symmetric non-singular (regular) black hole (RBH), which Bardeen was the first one to built in \cite{bardeen}. This and many other RBH models violate the strong energy condition which allows to break the singularity \cite{rbh}.  \par
 Cosmic strings are thought to be one-dimensional topological solitons that formed in the early universe during phase transitions \cite{kibble}. Locally, the string does not produce a gravitational field, but is globally conical. That means, outside the string, we see observable effects like light deflection \cite{aryal,lensingcosmic}. There are studies on finding cosmic strings that include this property, like the Capodimonte-Sternberg-Lens candidate no. 1 (CSL-1) \citep{csl1,csl2}. Another study is to find gravitational wave bursts produced by cosmic strings \citep{gw}. All these studies were promising, however, they do not have an exact observation of cosmic strings.
 \par
In this paper, we study how the string affects the thermodynamics properties of the RBH, largely studied in \cite{maluf}. We shall observe that it has not that much of an effect after all. \par
We will use the natural units throughout this paper, that is $G= \hbar =c=k_B =1$. 
\section{Regular black holes and cosmic strings}
\paragraph*{}
Since regular black holes have metrics with spherical symmetry, one can build the following metric in the $(t,r,\theta,\phi)$ coordinates 
\begin{equation}
ds^2=-f(r)dt^2+\frac{dr^2}{f(r)}+r^2(d\theta^2+{\alpha}^2sin^2{\theta}d\phi^2), \label{metric}
\end{equation}
\paragraph*{}
where cosmic string parameter is $\alpha=1-4\mu$, and $\mu$ is the mass per unit length of the string. The metric term $f(r)$ is
 \begin{equation}
 f(r)=1-{\frac{2m(r)}{r}}. \label{2}
\end{equation}  
\paragraph*{}
The mass function \cite{mass} is given by
\begin{equation}
m(r)={\frac{M_0}{{\left[1+{\left(\frac{r_0}{r}\right)}^q\right]}^{\frac{p}{q}}}}, \label{mass}
\end{equation}
where $M_0$ and $r_0$ are mass and length parameters, respectively. For an asymptotic flat spacetime, $p$ and $q$ are positive integers \cite{mass}. For $p=3$ and $q=2$, and $p=q=3$, the Bardeen and Hayward regular black holes are obtained, respectively. It is required for $p$ to be equal to 3, because in the limits of small $r$ of the mass function in Eq. \ref{mass}, we have a de Sitter core; firstly shown in \citep{core}, and improved in the 90s, see \citep{core2}.
 If $r_0<M_0$, two solutions, $r=r_{\pm}$, arise. Where $r_-$ is the inner, and $r_+$ is the outer horizon. The outer horizon is located at $r_+\:{\approx}\: 2m(r)$, as seen in Eq. \ref{2}.
\paragraph*{} 

\section{Temperature}
\paragraph*{}
We will derive the temperature in two ways; with the surface gravity $\kappa$, and with the tunneling effect.
\paragraph*{}
The surface gravity is 
\begin{equation}
\kappa=\left.\frac{1}{2}\frac{df(r)}{dr}\right|_{r_+}.
\end{equation}
In 1974, Hawking discovered that black holes emit radiation \cite{hawking}, therefore have a (Hawking) temperature given by 
\begin{equation}
T_{\kappa}=\frac{\kappa}{2\pi}.
\end{equation}
By using the metric in Eq. \ref{metric}, the mass function in Eq. \ref{mass}, and $M_0$, which can be found by $f(r_+)=0$, we can easily show that
\begin{equation}
f'(r)=\frac{2M_0{\left(1-\left(\frac{r_0}{r}\right)^q\right)}^{{-\frac{p}{q}-1}}\left[1-\left(p-1\right){\left(\frac{r_0}{r}\right)}^q\right]}{r^2},
\end{equation}
\begin{equation}
\kappa=\left.\frac{f'(r)}{2}\right|_{r=r_+}=\frac{M_0\left[1-\left(p-1\right){\left(\frac{r_0}{r_+}\right)}^q\right]{\left[1+\left(\frac{r_0}{r_+}\right)^q\right]}^{{-\frac{p}{q}-1}}}{r_+^2},
\end{equation}
\begin{equation}M_0=\frac{r_+}{2}\left[1+\left(\frac{r_0}{r_+}\right)^q\right]^{\frac{3}{q}}, 
\end{equation}
and therefore

\begin{equation} T_\kappa=\frac{1}{4\pi r_+}\left[1-2{\left(\frac{r_0}{r_+}\right)}^q\right] {\left[1+\left(\frac{r_0}{r_+}\right)^q\right]}^{{-1}}. 
\end{equation} \par

Hawking Radiation was largely studied in \cite{Sakalli:2014sea}, \cite{Sakalli:2015nza} and \cite{Sakalli:2016cbo}.

\paragraph*{}
The second way to calculate the temperature was the tunneling effect. Since only the radial trajectories are of interest, near horizon our metric can be studied as
\begin{equation}
ds^2=f(r)dt^2+\frac{dr^2}{f(r)}.
\end{equation}
 
 The Klein-Gordon equation for a scalar field $\phi$ and with mass $m_\phi$ is
 \begin{equation}
 \hbar^2g^{\mu\nu}\nabla_\mu\nabla_\nu\phi- m_\phi^2\phi=0. \label{11}
 \end{equation}
where $g^{\mu\nu}\nabla_\mu\nabla_\nu$ is known as the D'Alembert operator, $\Box$. The D'Alembertian is defined as, 
 \begin{equation} \Box=\frac{1}{\sqrt{|g|}}\partial_i \sqrt{|g|}g^{ij}\partial_j, \label{box}
 \end{equation}
where  $\:g=det(g_{ij}); $ with $\:i,j=\mu,\nu $.
We can write the metric tensor $g_{\mu\nu}$ as
 $$ g_{\mu\nu}=\begin{pmatrix} -f(r)&0\\0&f(r)^{-1}\end{pmatrix}, $$ 
$$\hspace*{-0.45cm}\Rightarrow g^{\mu\nu}=\begin{pmatrix}-f(r)^{-1}&0\\0&f(r)\end{pmatrix}.$$ 
Now we can calculate the D'Alembert operator. We can clearly see that $g=det(g_{\mu\nu})=-1$.
Using equation in Eq. \ref{box},

\begin{equation}
\Box\phi=\frac{1}{\sqrt{|-1|}}\partial_\mu \sqrt{|-1|}g^{\mu\mu}\partial_\mu+ \frac{1}{\sqrt{|-1|}}\partial_\nu \sqrt{|-1|}g^{\nu\nu}\partial_\nu,
\end{equation} where the other terms are equal to zero. Therefore,

\begin{equation}
 \Box\phi=-f(r)^{-1}\partial^2_t\phi+\partial_rf(r)\partial_r\phi+f(r)\partial^2_r\phi.
\end{equation} 

Put this in Eq. \ref{11}, 
\begin{equation}
\hbar^2\left[f(r)^{-1}\partial^2_t\phi+\partial_rf(r)\partial_r\phi+f(r)\partial^2_r\phi\right]-m^2_\phi\phi=0.
\end{equation} 

 We divide everything by $\hbar^2$ and multiply by $f(r)$ and get
\begin{equation}
\partial^2_t\phi+f(r)\partial_rf(r)\partial_r\phi+f(r)^2\partial^2_r\phi-\frac{{m^2_\phi}}{\hbar^2}f(r)\phi=0. 
\end{equation} We can write the $f(r)\partial_rf(r)\partial_r\phi $ part as $\frac{1}{2}\partial_rf(r)^2\partial_r\phi$. So the final form is,

\begin{equation} -\partial^2_t\phi+f(r)^2\partial^2_r\phi+\frac{1}{2}\partial_rf(r)^2\partial_r\phi-\frac{{m^2_\phi}}{\hbar^2}f(r)\phi=0. \label{17}
\end{equation}
To solve this equation, we use the WKB method, which has the ansatz solution given as
 \begin{equation}
 \phi(t,r)=exp\left[-\frac{i}{\hbar}I(t,r)\right] .
 \end{equation}

We put this in Eq. \ref{17};
\begin{equation}
-\partial^2_te^{-\frac{i}{\hbar}I\left(t.r\right)}+f^2\partial^2_re^{-\frac{i}{\hbar}I\left(t.r\right)}+\frac{1}{2}\partial_rf^2\partial_re^{-\frac{i}{\hbar}I\left(t.r\right)}-\frac{{m_{\phi}}^2}{\hbar^2}fe^{-\frac{i}{\hbar}I\left(t.r\right)}=0,
\end{equation} 

\begin{equation}
\frac{i}{\hbar}\partial^2_tI+\frac{1}{\hbar^2}\left(\partial_tI\right)^2-\frac{i}{\hbar}f^2\partial^2_rI-\frac{1}{\hbar^2}f^2\left(\partial_rI\right)^2-\frac{i}{\hbar}\frac{1}{2}\partial_rf^2\partial_rI-\frac{m_{\phi}^2}{\hbar^2}f=0.
\end{equation}

Multiply with $\hbar^2$,

\begin{equation}
 \hbar\partial^2_tI+\left(\partial_tI\right)^2-i\hbar f^2\partial^2_rI-f^2\left(\partial_rI\right)^2-i\hbar\frac{1}{2}\partial_rf^2\partial_rI-m_{\phi}^2f=0 .
\end{equation}

Lowest order of $\hbar$ yields the following Hamilton-Jacobi equation
\begin{equation}
\left(\partial_tI\right)^2-f(r)^2 \left(\partial_rI\right)^2-{m_{\phi}}^2f(r)=0,
 \end{equation}

and the split action form is
\begin{equation}
I(t,r)=-Et+W(r).
\end{equation}
$W(r)$ is the spatial part of the action, and is found to be
\begin{equation}
W_{\pm}=\pm\int \frac{dr}{f(r)}\sqrt{E^2-m_{\phi}^2f(r)},
\end{equation}

where $\pm$ corresponds to the outgoing and ingoing particles, respectively. Our focus for the Hawking radiation is the classically forbidden solutions $W_+(r)$, which cross the event horizon $r_+$.
\paragraph*{}
The coordinate invariant proper spatial distance is \cite{spatial},

\begin{equation}
d\sigma^2=\frac{dr^2}{f(r)}.
\end{equation}

The approximation of $f(r)$ near the horizon $r_+$ gives

\begin{equation}
f(r)=f'(r_+)(r-r_+),
\end{equation}

thus

\begin{equation}
\sigma=2\frac{\sqrt{r-r_+}}{\sqrt{f'(r_+)}},
\end{equation}
where $0<\sigma<\infty$.
\paragraph*{}
Then $W_+(r)$  reads

\begin{equation}
W_+(r)=\frac{2}{f'(r_+)}\int{\frac{d\sigma}{\sigma}\sqrt{E^2-\frac{\sigma^2}{4}m_\phi^2f'(r_+)^2}}=\frac{2{\pi}iE}{f'(r_+)}+real \:contribution.
\end{equation}
So the tunneling probability is given as

\begin{equation}
\Gamma\approx e^{-2ImI}=e^{-\frac{4{\pi}E}{f'(r_+)}}. \label{tunneling}
\end{equation}
 If we approach Eq. \ref{tunneling} like the Boltzmann factor $e^{-{E}/{T}}$, we get the Hawking temperature as
 
 \begin{equation}
 T_t=\frac{f'(r_+)}{4\pi}=T_\kappa
 \end{equation}
which is the same temperature as we found from the surface gravity, $\kappa$.
\paragraph*{}
As we can see, the temperature we have found is not different from the one in \citep{maluf}. Thus, the results are the same: we obtain the Schwarzschild temperature by setting $r_0=0$, $M_0=M$ and $r_+$ to the Schwarzschild radius, that is $T_{Sch}=1/4 \pi r_+$. And, with a little computation we see that (with or without cosmic strings, in this context) RBHs are colder than the Schwarzschild black holes.

\paragraph*{}

\section{Entropy}
\paragraph*{}

Having $r_+\approx 2m\left(r \right)$ as the position of event horizon, with the aid of Eq. \ref{metric}, we have the following horizon area

\begin{equation}
A=4\pi \alpha r_+^2=16\pi \alpha m^2. \label{31}
\end{equation} 

Since the area is $\alpha$-dependent, we have to follow a method given in \cite{aryal} to see the relation between $S$ and $A$.
\paragraph*{}
The following entropy relation is used,
\begin{equation}
dS=\frac{dE}{T}, \label{32}
\end{equation}
where $E$ is the measured energy of the black hole by an observer at infinity. Since with the string the spacetime is no longer asymptotically Minkowskian (flat), $E\neq M$.


\paragraph*{}
 Let $T_{\mu \nu}$ be the stress tensor for some matter field propagating on the spacetime, representing the Hawking radiation or classical matter that is thrown into the black hole. Also, let $\xi^{\mu}=(1,0,0,0)$ be the timelike Killing vector for our metric in Eq. \ref{metric}; then $\xi ^{\mu}T_{\mu \nu}$ is a covariantly conserved vector current and the rate of flow of energy in or out of the black hole may be written as

\begin{equation}
\dot{E}= \oint \xi^{\mu}T_{\mu \nu}d \Sigma ^{{\nu}},
\end{equation}
where the surface integral is taken over the horizon. The metric for a black hole given in Eq. \ref{metric} with slowly changing mass 
 \begin{equation}
m(r)=m(r,t)=m_0 + \dot{m}t,
\end{equation} 
where $m_0$ and $\dot{m}$ are constants. The Einstein tensor for this metric is

\begin{equation}
G_{\mu \nu}=G_{\mu \nu}^{(0)}+G_{\mu \nu}^{(1)}+\mathcal{O}\left(\dot{m}^2\right),
\end{equation}
where 
\begin{align*}
G_{\mu \nu}^{(0)}&=\Lambda g_{\mu \nu}, \\
G_{\mu \nu}^{(1)}&=R_{\mu \nu}-\frac{1}{2}Rg_{\mu \nu};
\end{align*} $R_{\mu\nu}$ and $R$ are the Ricci tensor and Ricci scalar, respectively, and $\Lambda$ is the cosmological constant \citep{ricci}. $G_{\mu \nu}^{(0)}=0$ and we neglect terms $\mathcal{O}\left(\dot{m}^2\right)$. The Einstein equations, $G_{\mu \nu}=8\pi T_{\mu \nu}$ then lead to

\begin{equation}
\begin{split}
\dot{E}&=\frac{1}{8\pi}\int_{r=r_+}G_{\mu \nu}\xi^{\mu}d\sum\nolimits^{\nu} \\[13pt] 
&=\frac{1}{8\pi} \int_{r=r_+} \frac{2\dot{m}}{r^2}\sqrt{g_{\theta \theta}g_{\phi \phi}}d\theta d\phi\\[13pt]
&=\frac{1}{8\pi}\frac{2\alpha \dot{m}}{r^2}\int_{r=r_+}r^2sin\theta d\theta d\phi \\[5pt]
\dot{E}&=\alpha \dot{m}.
\end{split}
\end{equation}

 As we can see, this solution confirms that the energy at infinity $E$ and the mass parameter $m$ are not identical. 
 \paragraph*{}
Now, if we use the new relation $dE=\alpha dm$, and the Hawking temperature $T=\frac{1}{8\pi m}$, in Eq. \ref{32},

\begin{equation}
\begin{split}
dS &= \alpha \frac{dm}{dT} \\
S &= 8\pi \alpha \int mdm \\
&=4\pi \alpha m^2. 
\end{split}
\end{equation}

Comparing it to Eq. \ref{31}, we clearly see that
\begin{equation}
S=\frac{1}{4}A,
\end{equation}
 which shows that the relation between $S$ and $A$ is not different with a string present from the relation that Bekenstein \cite{bekenstein} and Hawking \cite{hawking} stated.




\section{conclusion}

Our purpose in this paper was to study the effects a cosmic string has on the thermodynamics properties of the RBH. We observed the string does not change the temperature of the RBH. The relation of entropy and horizon area is also seen unchanged with a string present. We had to use a method given in \cite{aryal} to find the entropy. 

\section{acknowledgement}

The author is grateful to Ali {\"O}vg{\"u}n for getting her to study this topic and answering all of her questions in patience.

\end{document}